\documentclass[aps,prd,preprint,superscriptaddress,tightenlines,nofootinbib]{revtex4}

\usepackage{graphicx}
\usepackage{dcolumn}
\usepackage{bm}

\newcommand{\Uos}{\Upsilon(1S)}
\newcommand{\Ufs}{\Upsilon(4S)}

\newcommand{\wIS}{ISR-ON}
\newcommand{\wNIS}{ISR-OFF}
   
\begin{document}

\preprint{CLEO CONF 06-4}   

\title{Improved Measurement of the Branching Fraction and Energy
Spectrum of  $\eta'$ from  $\Uos$ Decays}
\thanks{Submitted to the 33$^{\rm rd}$ International Conference on High Energy
Physics, July 26 - August 2, 2006, Moscow}

\author{O.~Aquines}
\author{Z.~Li}
\author{A.~Lopez}
\author{S.~Mehrabyan}
\author{H.~Mendez}
\author{J.~Ramirez}
\affiliation{University of Puerto Rico, Mayaguez, Puerto Rico 00681}
\author{G.~S.~Huang}
\author{D.~H.~Miller}
\author{V.~Pavlunin}
\author{B.~Sanghi}
\author{I.~P.~J.~Shipsey}
\author{B.~Xin}
\affiliation{Purdue University, West Lafayette, Indiana 47907}
% ...
\collaboration{CLEO Collaboration} 
\noaffiliation
\date{July 25, 2006}

\begin{abstract}
We present an improved  measurement of the $\eta'$ meson energy
spectrum  in $\Uos$ decays, using 1.2 $\mathrm{fb^{-1}}$ of data taken at
the $\Uos$ center-of-mass energy with the CLEO III detector. We
compare our results with models of $\eta '$  gluonic form factor
that have been suggested to explain the  unexpectedly large
 $B\to \eta 'X_s$ rate. Models based on perturbative QCD fail to fit
the data for large $\eta '$ energies, showing that Standard Model
strong interaction dynamics is not likely to provide an explanation for the
large rate of high energy $\eta '$ observed in $B$ decays.
\end{abstract}

\maketitle

\section{Introduction}
\begin{figure}[htbp]
\includegraphics*[width=3.0in]{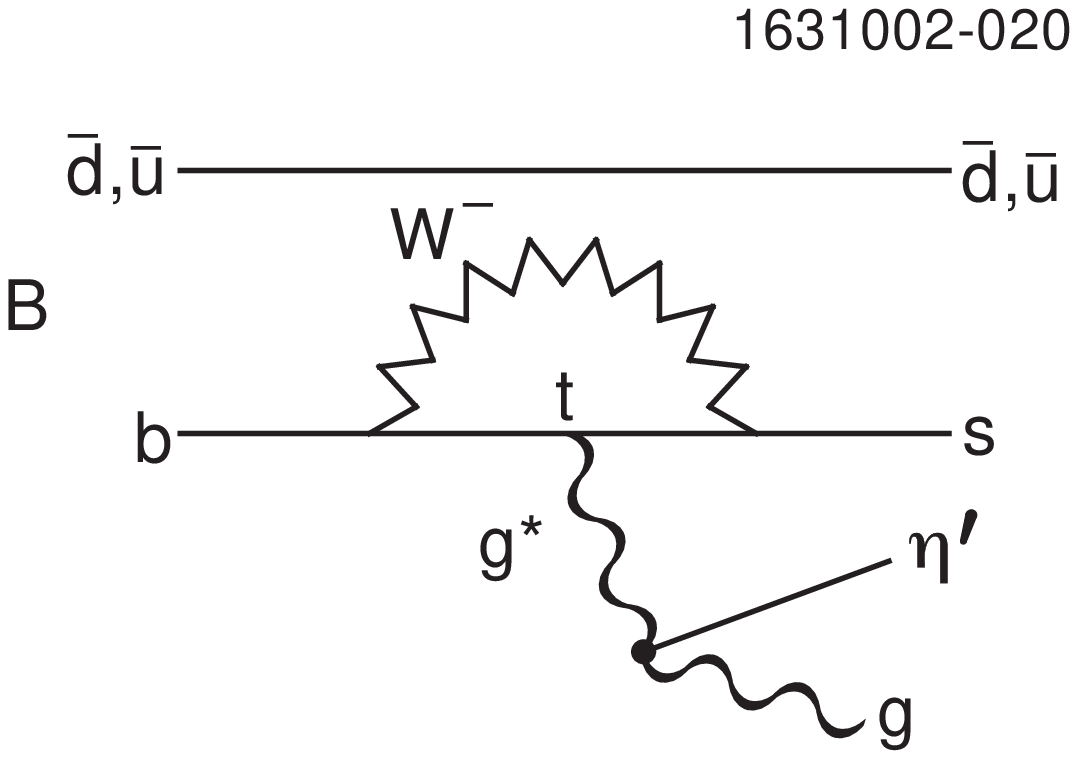}
\includegraphics*[width=3.0in]{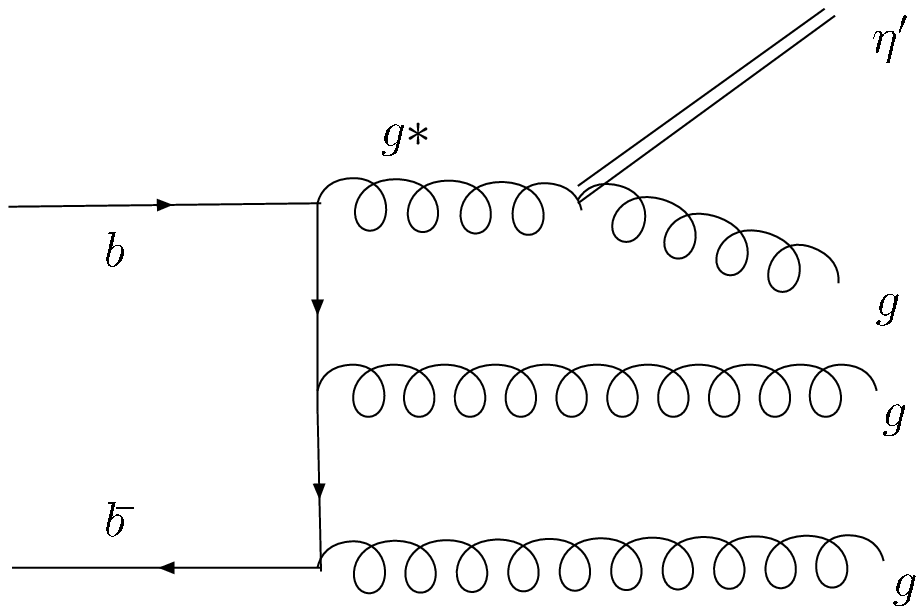}
\caption[A figure imported from a PostScript file] {Feynman Diagram
for $b\to s g$ (left) and $\Uos\to ggg^\star\to \eta ' X$ (right). }
    \label{etapvtx}
\end{figure}

A surprisingly large rate for $B \to \eta' X_s$, with high
momentum $\eta'$, ($p_{\eta '}$=2 - 2.7 GeV/c), was observed by
CLEO \cite{Browder,Ernst} and confirmed by BaBar \cite{babar}.
This result motivated intense theoretical activity because new
physics could account for such an enhancement. However, Standard
Model explanations have also been proposed. For example,  Atwood
and Soni \cite{Atwood:1997bn} argued that the observed excess is
due to an enhanced $b\to s g$ penguin diagram, complemented by a
strong $\eta'gg^{\star}$ coupling, induced by the gluonic content
of the $\eta '$ wave function. Fig.~\ref{etapvtx} (left) shows the
corresponding Feynman diagram. The high $q^2$ region of the
$g^\star g \eta '$ vertex function involved in this process  also
affects fast $\eta'$ production in $\Uos$ decay
\cite{Kagan02,Ali:2000ci}, whose relevant diagram is shown in
Figure~\ref{etapvtx} (right). Thus a precise measurement of the
$\eta '$ inclusive spectra from the process $\Uos\to ggg^\star \to
\eta' X$ provides valuable information towards our understanding
of important $B$ meson decays.

The effective vertex function $\eta' gg^\star$ can be written as
\cite{Atwood:1997bn}:
\begin{equation}
H(q^2)\delta^{ab}\varepsilon_{\alpha\beta\mu\nu}q^{\alpha}k^{\beta}\varepsilon^{\mu}_{1}
\varepsilon^{\nu}_{2},
\end{equation}
where $q$ is the $(g^\star)$ virtual gluon's four-momentum, $k$ is
the $(g)$ ``on-shell'' gluon's momentum ($k^2=0$), $a$, $b$
represent color indices,
$\varepsilon^{\mu}_{1},\varepsilon^{\nu}_{2}$ are the polarization
vectors of the two gluons, and $H(q^2)$ is the $g^\star g\eta'$
transition form factor. Different assumptions on its $q^2$
dependence have been proposed \cite{Atwood:1997bn,Hou:1997wy,
Kagan:1997,Ali:2000ci,pQCD,kroll}.

While ARGUS was the first experiment to study the inclusive $\eta '$
production at the $\Uos$ \cite{argus}, CLEO II \cite{jc} was the
first experiment to have sufficient statistics to measure inclusive
$\eta '$ production from the subprocess $\Uos\to ggg$. These data
ruled out a class of form factors characterized by a very weak $q^2$
dependence \cite{Atwood:1997bn,Hou:1997wy}. Intense theoretical
activity has followed \cite{Ali:2000ci,pQCD,kroll} to derive the
perturbative QCD form factors from models of the $\eta '$ wave
function. Attempts to use CLEO II data to constrain the model
parameters \cite{ali-extended} were not conclusive, due to the
limited statistics at the end point of the $\eta '$ spectrum.
Moreover, the order of the perturbative expansion necessary to
achieve a good representation of the data was not clearly defined
\cite{ali-extended}. Thus an improved measurement, based on a higher
statistics sample, is important to settle these issues. This work
reports a new measurement of the inclusive $\eta '$ spectrum from
the process $\Uos \to ggg^\star \to \eta ' X$ based on the largest
$\Uos$ sample presently available, more than a factor of 11 higher
than the previous study \cite{jc}. Thus the measured high energy
distribution function provides a much more stringent constraint.

\section{Data sample and analysis method}
We use 1.2 $\mathrm{fb^{-1}}$ of CLEO III data recorded at the $\Uos$
resonance, at 9.46 GeV center-of-mass energy, containing $21.2\times
10^6$ events and off-resonance continuum data collected at
center-of-mass energies of 10.54 GeV
(2.3 $\mathrm{fb}^{-1}$).

The CLEO III detector includes a high resolution tracking system
\cite{track}, a state of the art CsI electromagnetic calorimeter
\cite{csi}, and a Ring Imaging Cherenkov (RICH) hadron
identification system \cite{rich}. The CsI calorimeter measures the photon
energies with a resolution of 2.2\% at $E=1$ GeV and 5\% at $E$=100
MeV. The tracking system provides also charged particle
discrimination, through the measurement of the specific ionization
$dE/dx$.

We detect $\eta '$ mesons through the channel $\eta '\to\eta
\pi^+\pi^-$ and $\eta \to \gamma\gamma$. The branching fractions
for these processes are (44.5$\pm$1.4)\% and (39.38$\pm$0.26)\%
respectively. We identify single photons based on their shower
shape. The photon four-vectors are constrained to have invariant
mass equal to the nominal $\eta$ mass. Subsequently, $\eta$
candidates are combined with two oppositely charged tracks to form
an $\eta '$. Loose $\pi$ consistency criteria based on $dE/dx$
measurements are applied to the charged tracks.

The gluonic $\eta '$ production at the $\Uos$  is expected  to be
dominant only at very high $q^2$, or, equivalently, at high $\eta '$
scaled energy $Z$, where $Z$ is defined as
\begin{equation}
Z\equiv
\frac{E_{\eta '}}{E_{\text{beam}}}
=\frac{E_{\eta '}}{2M_{\Uos}},
\end{equation}
where $E_{\eta '}$ is the $\eta '$ energy and $E_{\text{beam}}$ is
the beam energy. Enhanced $\eta '$ production at high $Z$ would
indicate a large $\eta 'g^{\star} g$ coupling.

\begin{figure}[htbp]
\includegraphics*[width=6in]{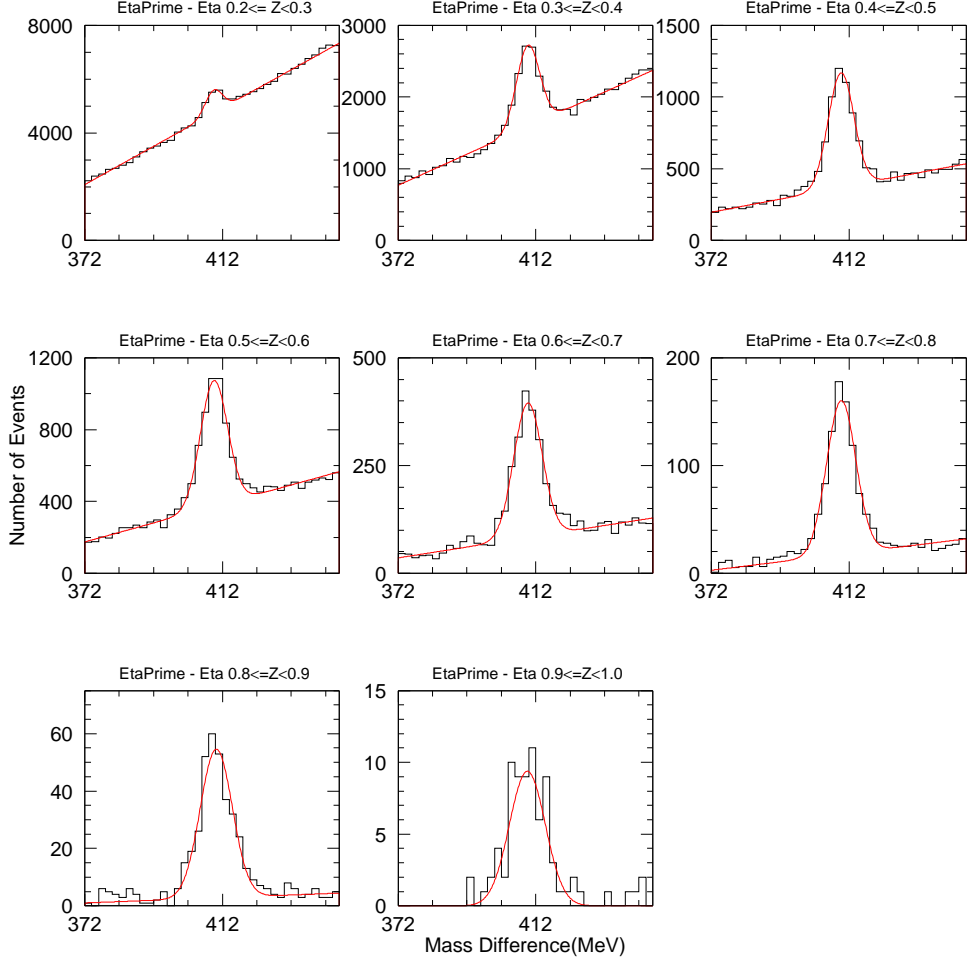}
 \caption{\label{fig:xm1sz} The difference of the $\eta\pi^+\pi^-$ and $\eta$ invariant masses spectra in
  different Z ranges reconstructed from $\Uos$ data,
  fit with a single Gaussian function for the signal and a first order
  polynomial for the background.}
\end{figure}

\begin{figure}[htbp]
\includegraphics*[width=6in]{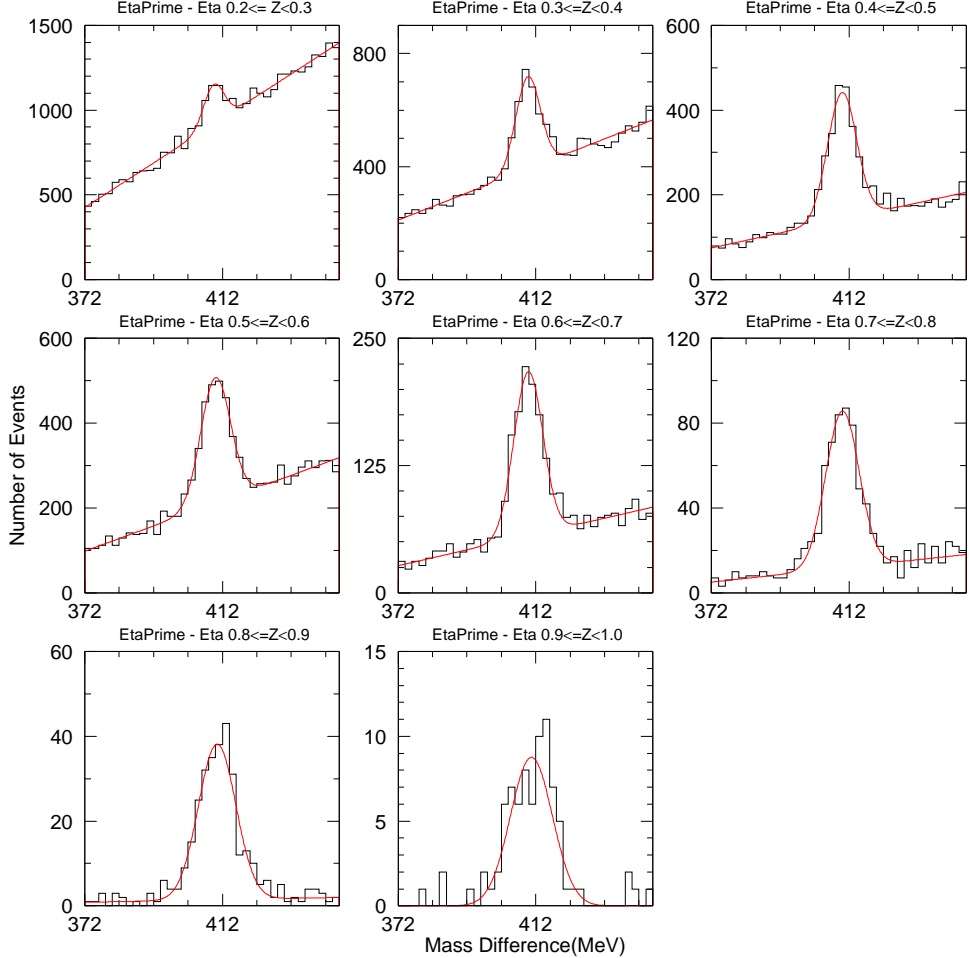}
 \caption{\label{fig:xm4sz} The difference of the $\eta\pi^+\pi^-$ and $\eta$ invariant masses spectra in
  different Z ranges reconstructed from continuum data taken at
  a center-of-mass energy of 10.54 GeV,
  fit with a Single Gaussian function for the signal and a first order
  polynomial for the background.}
\end{figure}

For low energy $\eta '$s, photons
coming from low energy $\pi ^0$s are a severe source of background.
Thus a $\pi ^0$ veto is applied for $Z < 0.5$,
whereby photon pairs that have an invariant mass consistent
(within 2.5 $\sigma$) of the nominal $\pi ^0$ mass are not
included as the candidate photons for $\eta$ reconstruction. We
consider only $\eta '$ with scaled energy $Z$ between 0.2 and 1 and
divide this range into eight equal bins. Fig.~\ref{fig:xm1sz}
demonstrates the extraction of the $\eta '$ yields in these bins for the $\Uos$
sample. Fig.~\ref{fig:xm4sz} shows the corresponding plots from
the continuum sample taken at a center-of-mass energy of 10.54
GeV. In order to derive the $\eta '$ signal yields,  we fit the
mass difference spectra $\Delta M_{\eta '\eta}\equiv
M(\pi^+\pi^-\eta)-M(\eta)$ in each $Z$ bin with a Gaussian
function representing the signal, and a polynomial background. The
Gaussian is used only to define a $\pm 2.5 \sigma$ signal
interval. Then the $\eta '$ yield in this interval is evaluated
counting events in the signal window, after subtracting the
background estimate obtained from the fit function. As the signal
is not described well by a single Gaussian function, this
procedure minimizes systematic uncertainties associated with the
choice of an alternative signal shape.


Information on the gluon coupling of the $\eta '$ can be drawn
only from the decay chain $\Uos \to g^{\star}gg\to \eta ' X$. Thus
we need to subtract both continuum $\eta '$ production and $\eta
'$ from the process $\Uos \to \gamma ^\star\to q\bar{q}$. The
latter component is estimated using
\begin{equation}
{\cal B}(\Uos\to q\bar{q})= R\cdot {\cal B}(\Uos\to \mu ^+\mu^-)=
(8.83\pm 0.25)\%,
\end{equation}
where $R$ is the ratio between the hadronic cross section $\gamma
^\star \to q\bar{q}$ and the di-muon cross section $\gamma ^\star
\to \mu ^+\mu^-$ at an energy close to 9.46 GeV. We use $R=3.56\pm
0.07$ \cite{ammar:1998} and ${\cal B}(\Uos\to \mu ^+\mu^-)= (2.48
\pm 0.05)$\% \cite{pdg06}.

\begin{figure}[htbp]
\includegraphics*[width=3.75in]{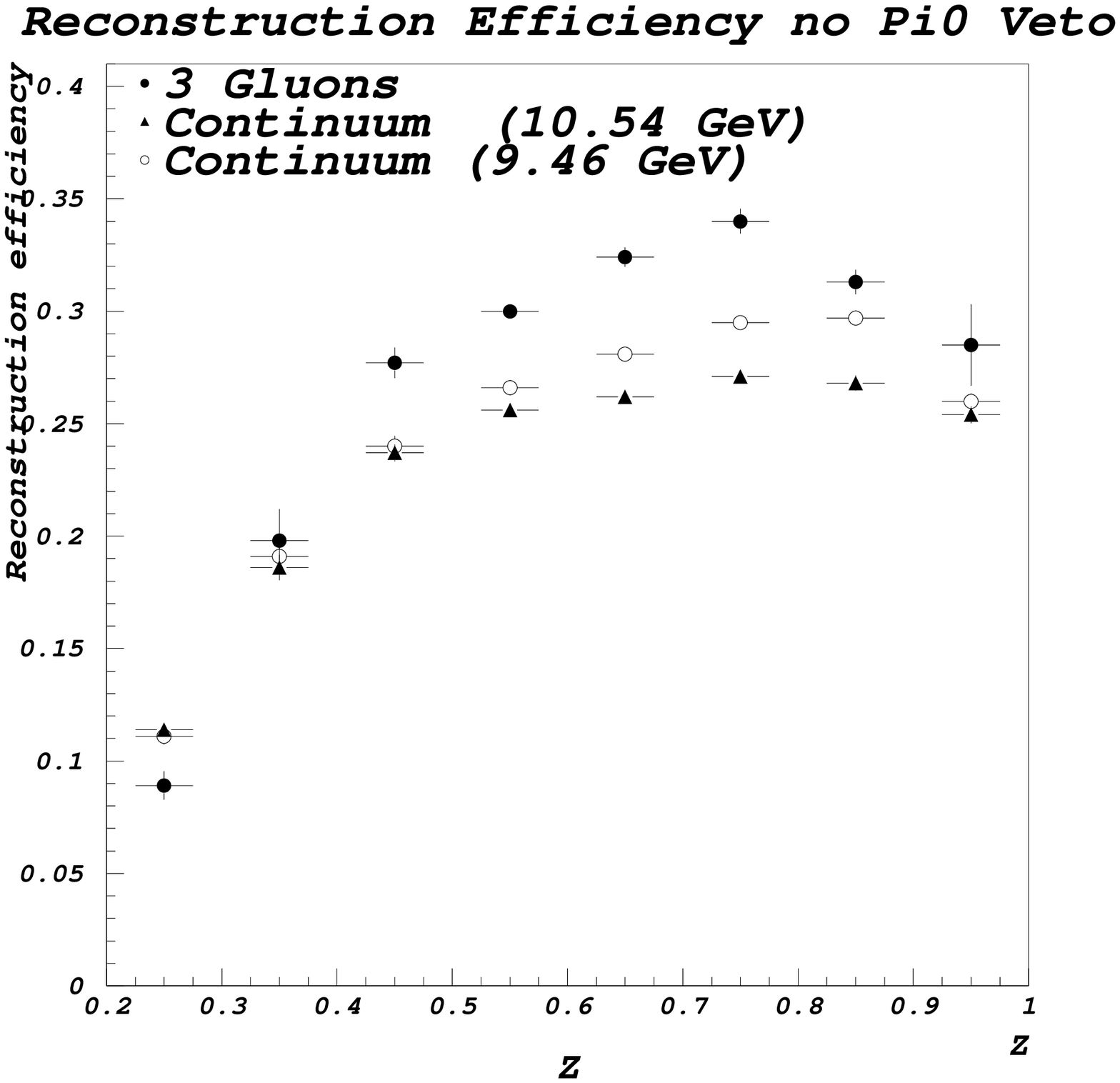}
\includegraphics*[width=3.75in]{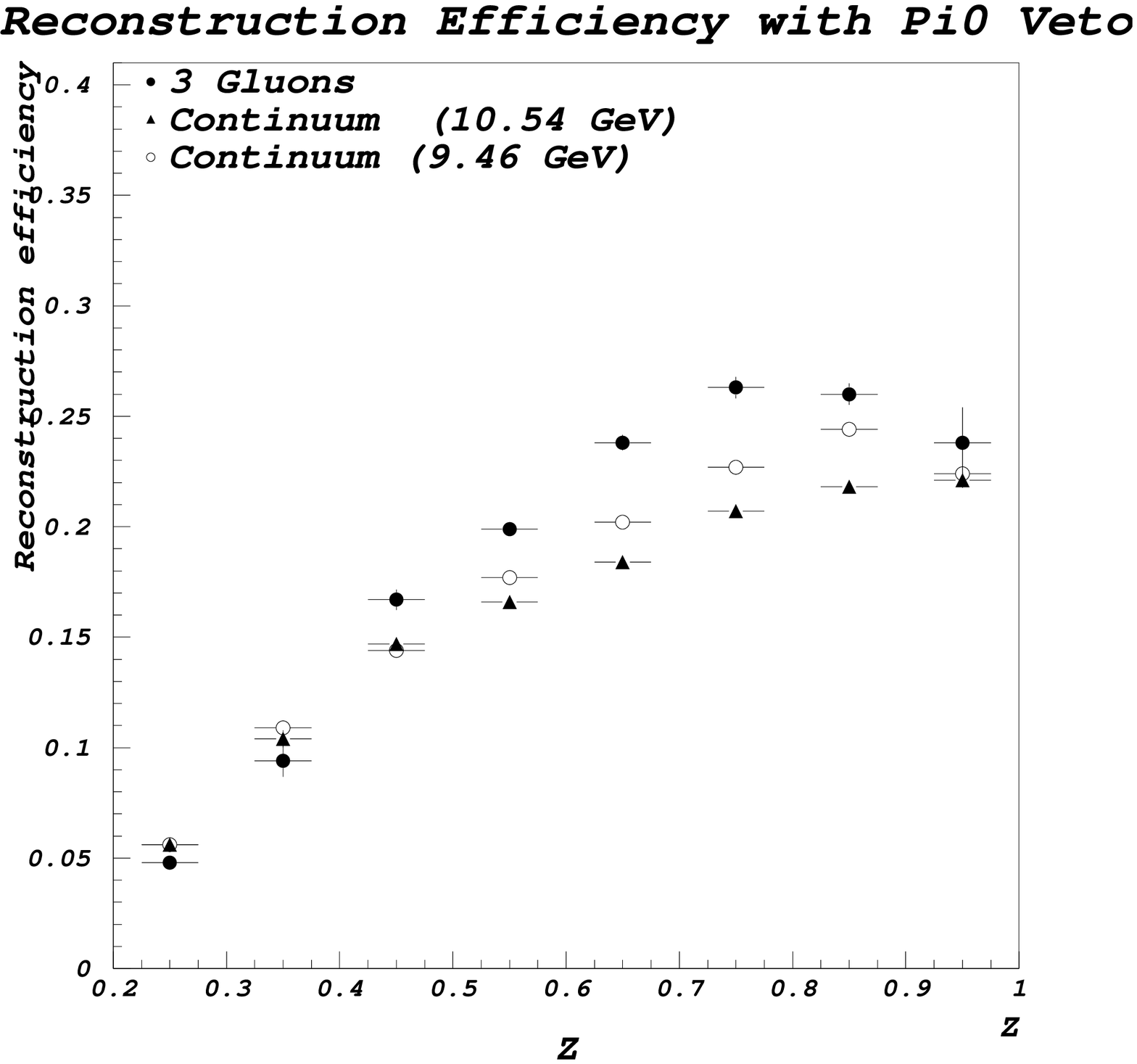}
  \caption{\label{fig:eff}The $\eta'$ reconstruction efficiencies as function of Z
 for different MC samples with no $\pi^o$ veto, and with $\pi^0$ veto in
 photon selection. The $\pi^0$ veto was applied in this analysis for $Z<0.5$.}
\end{figure}

The two dominant components of the $\eta '$ spectrum have different
topologies: $\Uos \to ggg$ produces a spherical event topology,
whereas $q\bar{q}$ processes are more jet-like. This difference
affects the corresponding reconstruction efficiencies.
Fig.~\ref{fig:eff} shows the efficiencies obtained for the two event
topologies with  CLEO III Monte Carlo studies.  We use GEANT-based
\cite{geant} Monte Carlo samples, including $\Uos$ and continuum
samples. In order to use the continuum sample taken at 10.54 GeV
center-of-mass energy for background subtraction, we develop a
``mapping function", to correct for the difference in phase space
and $Z$ range spanned in the two samples. The procedure is described
in detail in Ref.~\cite{jc}.  By comparing the $\eta '$ energy
distribution functions  for the Monte Carlo continuum samples at
center-of-mass energies equal to 9.46 and 10.54 GeV, we obtain the
mapping:
\begin{equation}\label{eq:Z}
Z_{9.46} = -0.215 \times 10^{-2} + 1.2238\ Z_{10.54} - 0.6879\
(Z_{10.54})^2
               + 0.8277\ (Z_{10.54})^3 - 0.3606\ (Z_{10.54})^4\ , \\
\end{equation}
where $Z_{9.46}$ is the $Z$ value used to subtract the continuum
contribution, as mapped from $Z_{10.54}$, the measured $Z$ in the
continuum data taken at the center-of-mass energy equal to 10.54
GeV.

The $\gamma gg/ggg$ cross section ratio is only about 3\%, thus we
make no attempt to subtract this term.

\begin{figure}[htbp]
\includegraphics*[width=5.in]{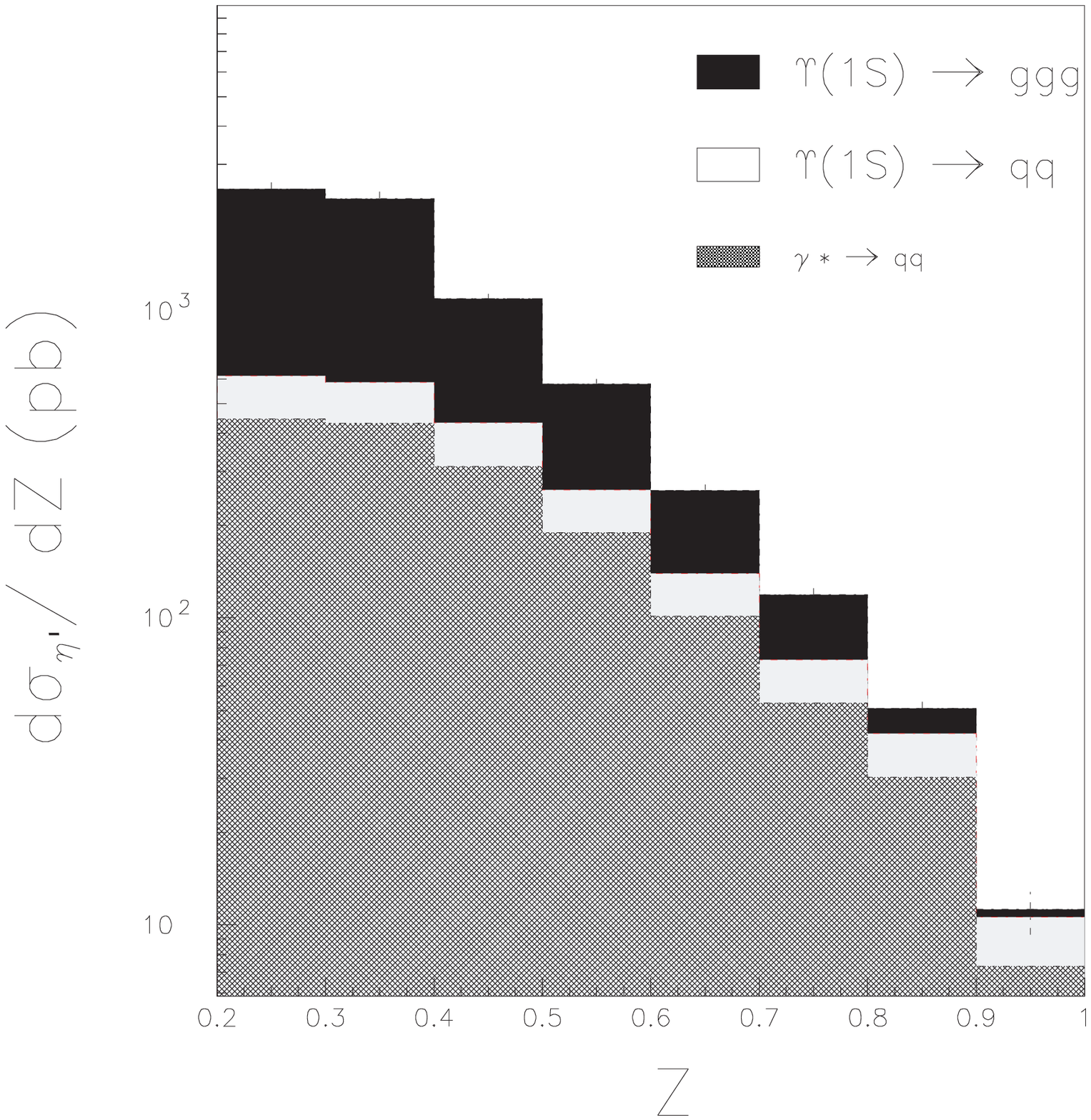}
\caption{\label{fig:etapcross}The differential cross sections
$d\sigma/dZ$ a) $\gamma ^\star\to q\bar{q}\to\eta'X$ (hatched), b) $\Uos\to
q\bar{q}\to\eta'X$ (white) and c) $\Uos\to ggg\to\eta'X$ (black).}
\end{figure}

Upon correcting the raw yields with the corresponding efficiencies,
the $\eta '$ spectrum from the 3 gluon decay of the $\Uos$ is
extracted using the relationship:
\begin{equation}
N(\Uos \to ggg)= N_{\mathrm{had}} - N(\gamma^\star\to q\bar{q})- N(\Uos\to
q\bar{q}),\label{eq:ggg}
\end{equation}
where $N_{\mathrm{had}}$ is the number of hadronic events in our
sample, and $N(\gamma^\star\to q\bar{q})$ is the number of
continuum events estimated from the sample taken at 10.54 GeV,
corrected for the luminosity difference between the resonance and
continuum data, and the $s$ dependence of the cross section for
the process $\gamma ^\star \to q\bar{q}$. The component
$N(\Uos)\to q\bar{q}$ is estimated from $N(\gamma^\star\to
q\bar{q})$
 using the relationship:
\begin{eqnarray}
N(\Uos \to q\bar{q})& = & N(\gamma^\star\to q\bar{q})\times R_{\mathrm{\wNIS / \wIS}} \times \frac{\sigma_{\Uos\to q\bar{q}}}{\sigma_{e^{+}e^{-}\to q\bar{q}}} \nonumber \\
            & = & N(\gamma^\star\to q\bar{q})\times R_{\mathrm{\wNIS / \wIS}} \times \frac{\sigma_{\Uos\to\mu^{+}\mu^{-}}}{
\sigma_{e^{+}e^{-}\to\mu^{+}\mu^{-}}}, \label{eq:fqq}
\end{eqnarray}
where $R_{\mathrm{\wNIS / \wIS}}$ accounts for the difference between the
$\Uos\to q\bar{q}\to \eta ' X$ and the $\gamma ^\star\to q\bar{q}\to
\eta ' X$ spectra due to initial state radiation  (ISR) effects,
estimated using Monte Carlo continuum samples with and without ISR
simulation, and $\sigma(\Uos\to q\bar{q})/\sigma(e^{+}e^{-}\to
q\bar{q})$ accounts for the relative cross section of these two
processes. The correction factor $R_{\mathrm{\wNIS / \wIS}}$ differs from 1
by a few percent at low $Z$ and as much as 25\% at the end point of
the $\eta '$ energy. The cross sections used are
$\sigma(\Uos\to\mu^{+}\mu^{-})=0.502\pm 0.010 $ nb \cite{pdg06} and
$\sigma(e^{+}e^{-}\to\mu^{+}\mu^{-})$ is 1.38 nb \cite{fpair}.
Fig.~\ref{fig:etapcross} shows the differential cross sections
$d\sigma/dZ$ for the processes $\Uos\to ggg$, $\Uos \to q\bar{q}$,
and $\gamma ^\star \to q\bar{q}$.

\begin{figure}[htbp]
\includegraphics*[width=3.in]{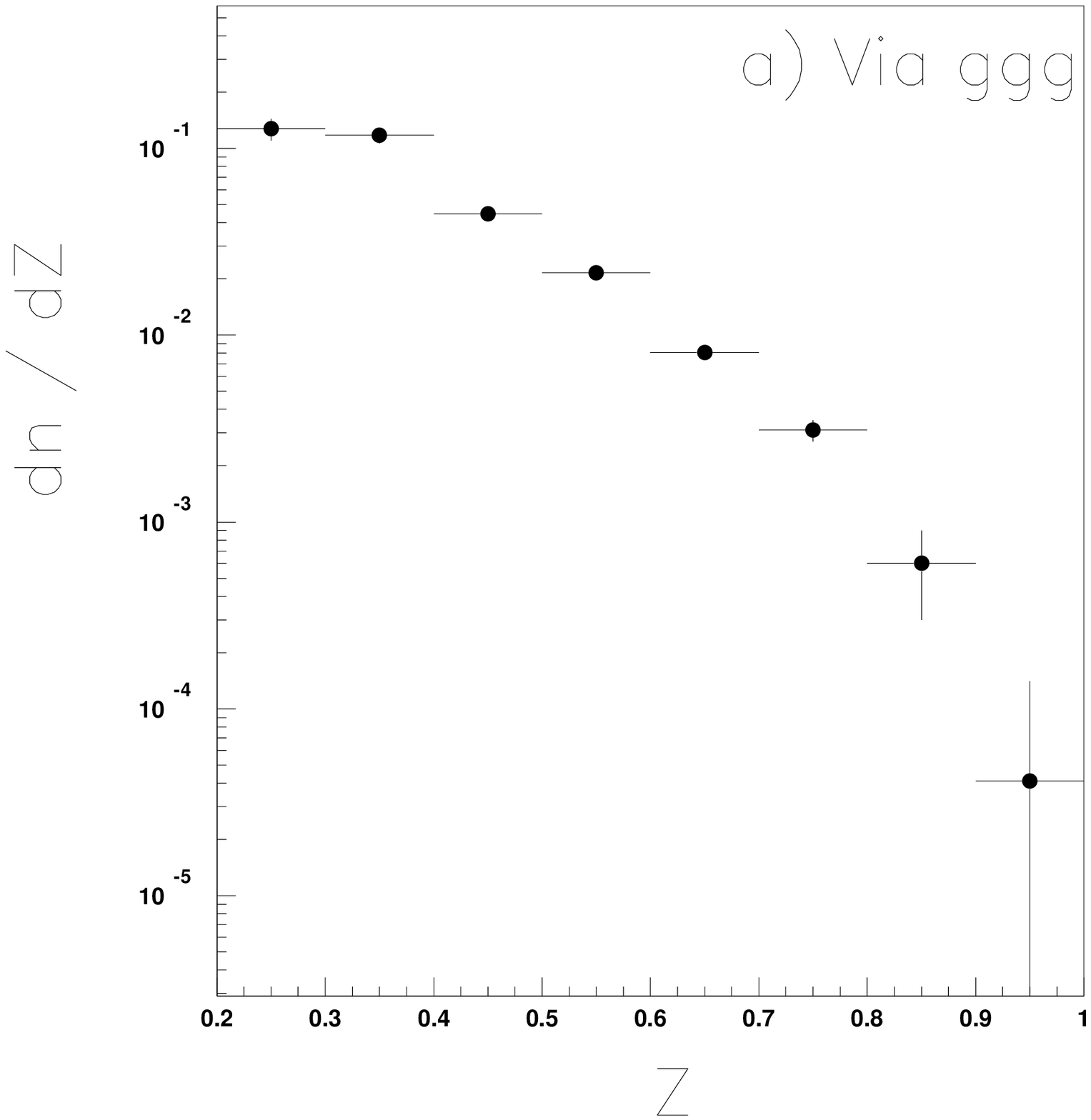}
\includegraphics*[width=3.in]{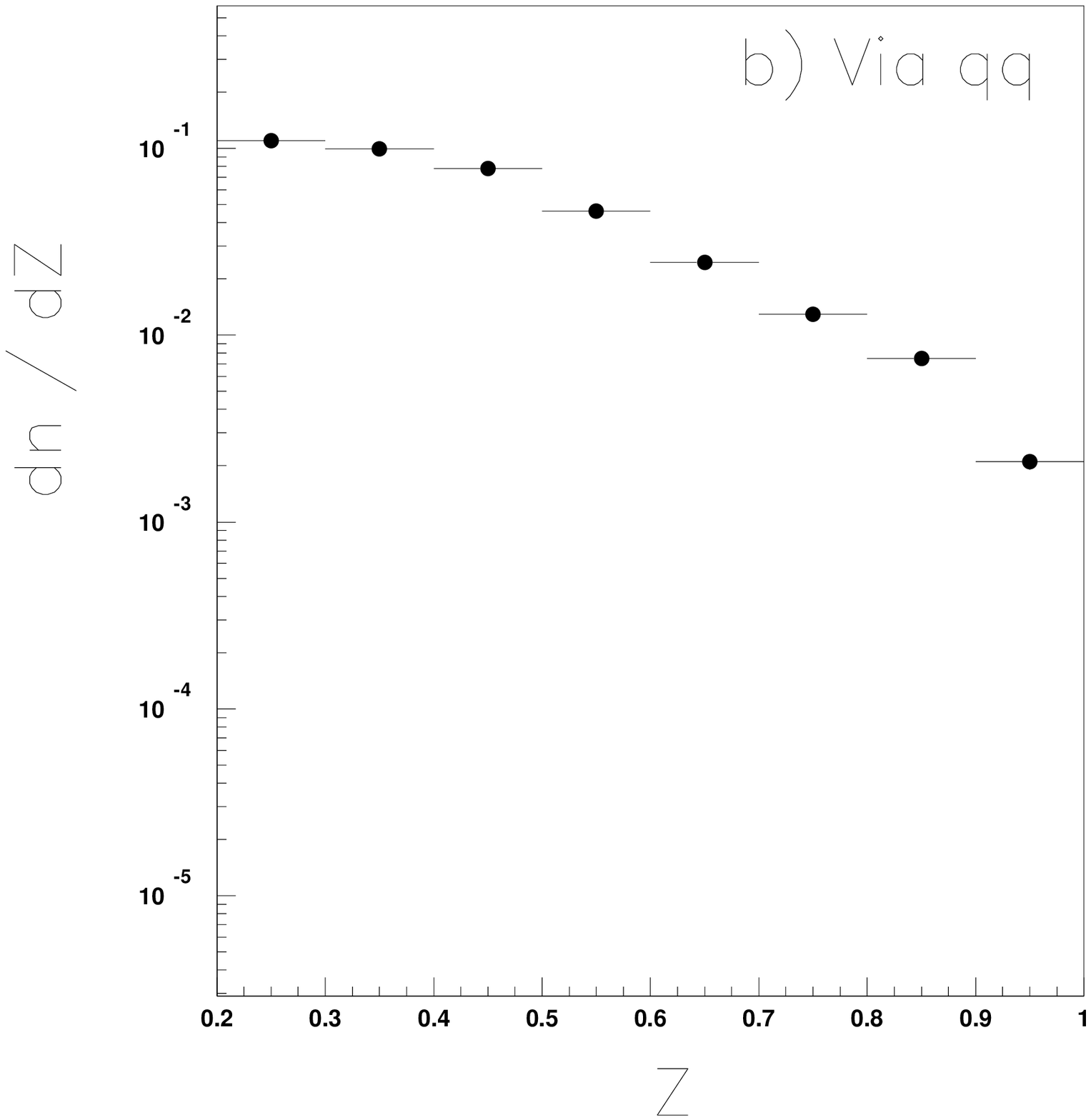}
\includegraphics*[width=3.in]{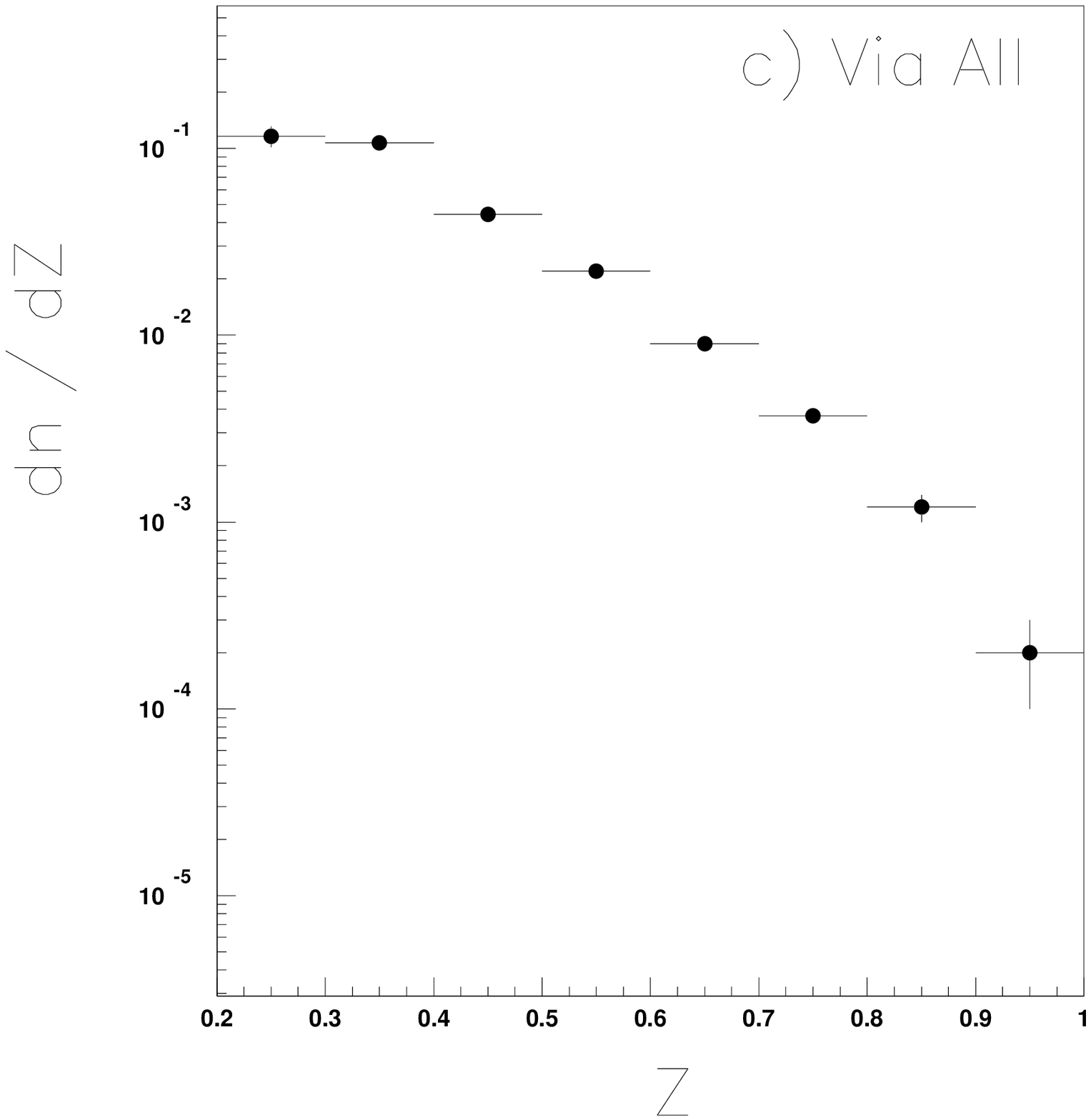}
 \caption{\label{fig:zspect}The energy distribution function $dn/dZ$ as defined in
  context for a) $\Uos\to ggg\to\eta'X$, b) $\Uos\to q\bar{q}\to\eta'X$, and c)$\Uos\to\eta'X$.}
\end{figure}

Theoretical predictions give the energy distribution function
$$
\frac{dn}{dZ} \equiv \frac{1}{N(\Uos \to ggg)}
                 \times \frac{dN(\Uos\to ggg)}{dZ}; 
$$
we obtain
the corresponding experimental quantity by dividing by the total
number of $N(\Uos \to ggg)$, estimated by applying Eq.~\ref{eq:ggg}
without any $Z$ restriction. Fig.~\ref{fig:zspect}.a) shows the
$\Uos \to ggg\to \eta ' X$ energy distribution function, whereas
Fig.~\ref{fig:zspect}.b) and c) show the corresponding distributions
for $\Uos \to q\bar{q}\to \eta ' X$, normalized with respect of the
total number of $\Uos\to q\bar{q}$ and $\Uos \to \eta ' X$,
normalized with respect to the total number of $\Uos$.

The inclusive $\eta '$ production at the $\Uos$ is expected to be dominated
by the transition $\Uos\to ggg^\star\to \eta 'X$ only at high $\eta '$
energy. The energy at which this occurs cannot be predicted from first
principle: an empirical criterion is the $\chi ^2$ of the theory fit to the
data. For example, a numerical analysis of the CLEO II data
\cite{ali-extended} obtained a $\chi ^2$ of 2.4 for three degrees of freedom,
using the 3 experimental points at $Z\ge 0.7$, and $\approx$ 24 for 4 degrees
of freedom using the 4 points at $Z\ge 0.6$. Thus Ali and Parkhomenko
concluded that the $Z$ region
likely to be dominated by the $\Uos\to ggg ^\star\to \eta ' X$
starts at $Z\ge 0.7$. Thus, we quote separately global branching
fractions for $\Uos\to \eta 'X$ and the corresponding results for
$Z\ge 0.7$.

Table~\ref{tab:error} summarizes the dominant components of the
systematic uncertainties. The systematic errors on the branching
fractions from $\eta '$ are $\pm 8.1\%$ for $q\bar{q}$, $\pm
9.1\%$ for the $ggg \to \eta ' X$ for $Z>0.7$ and $\pm 7.2\%$ for
all other branching fractions.

\begin{table}[ht]
\begin{center}
\begin{tabular}{|l|c|c|c|}
\hline \hline Sources     & $ggg$ Sample ($Z>0.7$) & $q\bar{q}$
Sample & All others \\ \hline
Reconstruction efficiency of $\pi^\pm$  & 2 & 2 & 2\\
Reconstruction efficiency of $\eta$     &   5 &   5 &   5 \\
Number of $\eta'$ from fit              &   1 &   1 &   1 \\
Total number of $\Uos$                  & 1 & 1 & 1 \\
${\cal B}(\eta'\to\pi^+\pi^-\eta)$      & 3.4 & 3.4 & 3.4 \\ \hline
${\cal B}(\Uos\to q\bar{q})$            &  -  & 3 & -   \\
Ratio of integrated luminosity         & 1.9 &  1  & -   \\
$\sigma_{\Uos\to\mu^+\mu^-},\sigma_{e^{+}e^{-}\to\mu^{+}\mu^{-}}$            & 0.7 &  1.3  & -   \\
$\pi^0$ veto                            &  -  & 1.7 & 0.4 \\
Z mapping                               &   6 &  3  & 3   \\ \hline
Total                                   &  9.1 & 8.1  & 7.2 \\
\hline\hline
\end{tabular}
\caption{\label{tab:error} The components of the systematic errors
(\%) affecting the branching fractions reported in this paper.}
\end{center}
\end{table}

Thus we obtain:
\begin{eqnarray}
n(\Uos\to (ggg) \to\eta'X)\equiv \frac{N( \Uos\to ggg
           \to\eta'X)}{N(\Uos\to ggg)}
           &=& (3.2 \pm 0.2 \pm 0.2) \%, \nonumber\\
n(\Uos\to(q\bar{q})\to\eta'X)\equiv \frac{ N(\Uos\to
           q\bar{q}\to\eta'X)}{N(\Uos\to q\bar{q})}
           &=& (3.8 \pm 0.2 \pm 0.3)\%, \nonumber\\
n(\Uos\to\eta'X)\equiv \frac{ N(\Uos\to\eta'X)}{N(\Uos)}
           &=& (3.0 \pm 0.2 \pm 0.2) \%. \label{eq:br}
\end{eqnarray}

The $\Uos\to\eta'X$ branching fractions at high momentum ($\rm
Z>0.7$) are measured to be:
\begin{eqnarray}
n(\Uos\to(ggg)\to\eta'X))_{Z>0.7}
           &=& 3.7\pm 0.5 \pm 0.3)\times 10^{-4}, \nonumber\\
n(\Uos\to(q\bar{q})\to\eta'X))_{Z>0.7}
           &=& (22.5\pm 1.2 \pm 1.8)\times 10^{-4}, \nonumber\\
n(\Uos\to\eta'X)_{Z>0.7}
           &=& (5.1 \pm 0.4 \pm 0.4)\times 10^{-4}.\label{eq:br7}
\end{eqnarray}
\section{Comparison with theory and conclusions}
A. Kagan \cite{Kagan02} used the ratio $R_{Z>0.7}$ defined as
\begin{equation}
R_{Z>0.7}\equiv\left[\frac{n_{th}}{n_{exp}}\right]_{Z>0.7},
\end{equation}
to obtain a first rough discrimination between radically different
$q^2$ dependence between the form factors. At the time that
Ref.~\cite{Kagan02} was published, the comparison was based on
90\% c.l. upper limits on the data. This test repeated with our
present data give values of $R_{Z>0.7}$ equal to 74 for a
representative slowly falling form factor \cite{Hou:1997wy}, 25
for the intermediate form factor studied by
Ref.~\cite{Kagan:1997}, and 2 for the perturbative QCD inspired
shape. Thus the last shape is the closest to the data, although
not providing a perfect match.

\begin{figure}[htbp]
\includegraphics*[width=3.75in]{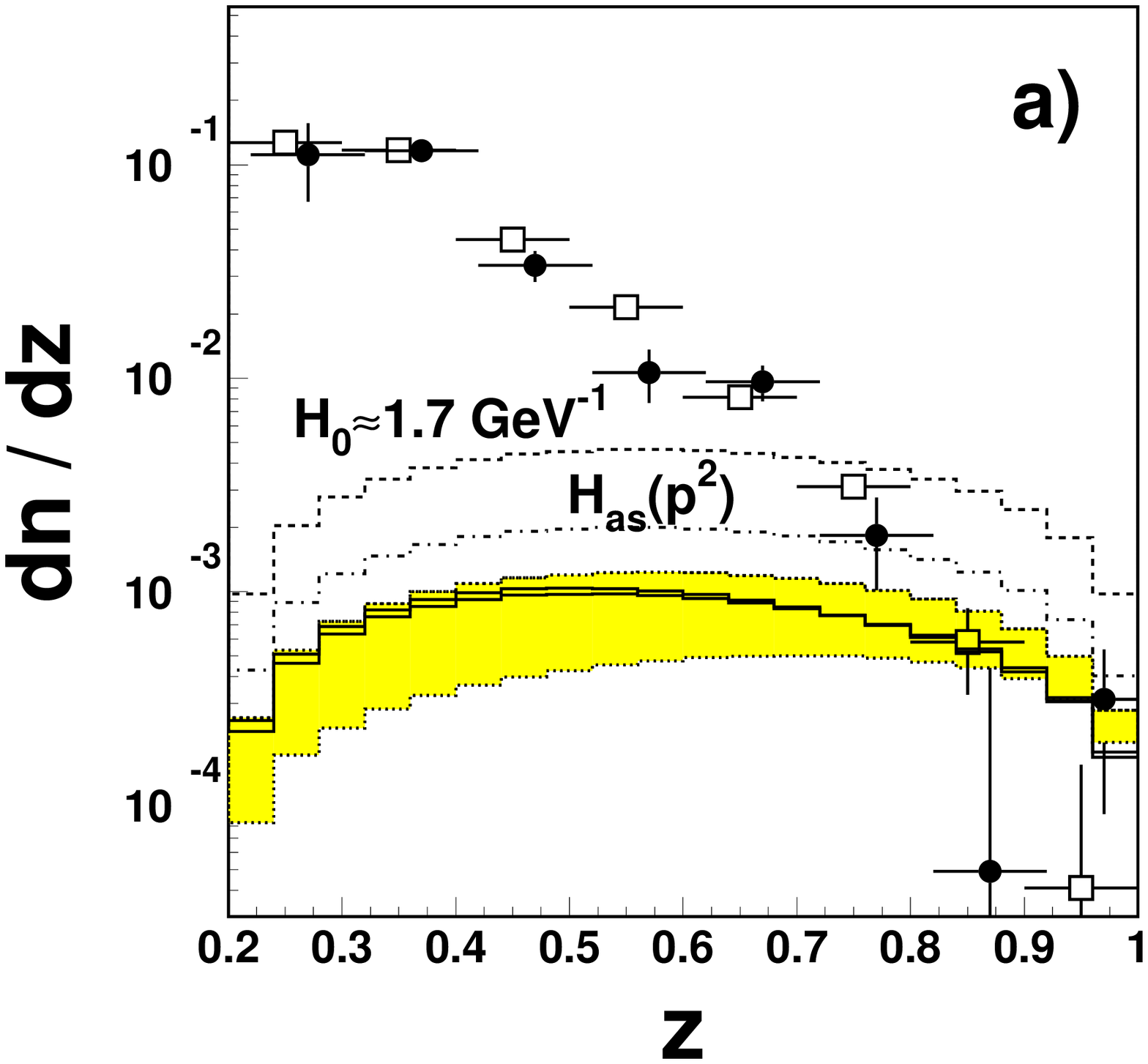}
\includegraphics*[width=3.75in]{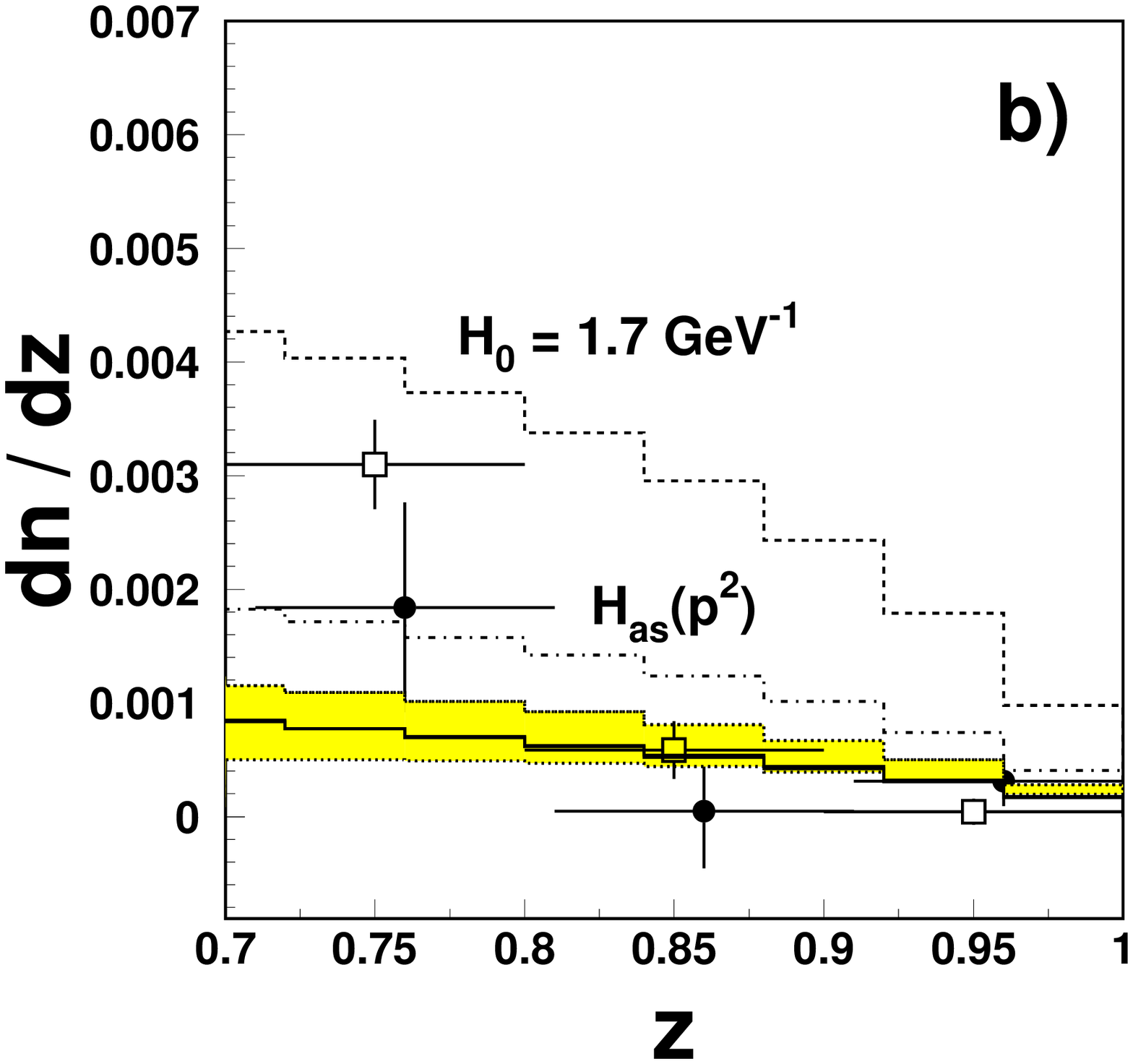}
    \caption[A figure imported from a PostScript file]{
    Energy spectrum of the $\eta '$-meson in the decay
    $\Uos\to\eta'X$: measured spectra: open squares correspond to  the data
presented in this paper, filled circles are the previously reported CLEO II 
data \cite{jc}:a) measured spectra b) $Z\ge 0.7$ region zoomed in
 to show the comparison with the theoretical
    predictions more clearly. The dashed curve corresponds to a constant
    value of the function $H(p^{2})=H_{0}\simeq 1.7GeV^{-1}$, and the
    dash-dotted curve ($H_{as}(p^{2})$) corresponds to the asymptotic from
    of the $\eta '$-meson LCDA \cite{jc} (i.e., $B_{2}^{(q)}=0$ and
    $B_{2}^{(q)}=0$). The spectrum with the Gegenbauer coefficients in
    the combined best-fit range of these parameters is shown in the
    shaded region. The solid curve 
    corresponds to the best fit values of the Gegenbauer coefficients
    from the analysis of the $\Uos\to\eta'X$ CLEO II data alone. The
    dotted points represent the CLEO II data \cite{ali-extended}
        and the boxed
    points represent the CLEO III data.
    }
    \label{fig:cleocomp}
\end{figure}

Different approaches have been taken to implement perturbative QCD
calculations. Their difference is translated into different
assumptions for the form factor $H(q^2)$,  Kagan and Petrov \cite{Kagan02} assume
$H(q^2)\approx\ \text{const} \approx 1.7\ \rm{GeV}^{-1}$, Ali and
Parkhomenko related $H(q^2)$ to     the expansion of the two
light-cone distribution amplitudes (LCDA) describing the quark and
gluon components of the $\eta '$ wave function.
Fig.~\ref{fig:cleocomp} shows the measured $dn/dZ$ distribution,
compared with three representative choices for $H(q^2)$.
$H(q^2)=H_0=1.7 GeV^{-1}$, $H_{as}$, based on the asymptotic form of
the $\eta '$ meson LCDA, and $H(q^2)$ corresponding to the spectrum with the
Gegenbauer coefficients from the best fit range obtained by Ali and
Parkhomenko \cite{ali-extended}, using the previous CLEO II data and
the constraints from the $\eta '-\gamma$ transitions \cite{kroll}.
Note that most of the discrepancy between theory and data occurs in
the $Z=0.7$ bin. In fact, the $\chi ^2$ for the fit of the new data
with this theoretical parametrization is 27. This may
imply that higher order terms in the QCD expansion need to be taken
into account, or that the $\Uos \to 3 g$ is not the dominant source
of $\eta '$, at least at a scaled energy as high as  $Z=0.7$.

In conclusion we have measured the energy spectra of the $\eta '$
meson in the decay $\Uos \to \eta ' X$. Our results are not very
well described by existing models based on strong gluonic coupling of the
$\eta '$. Thus the observed $B\to \eta ' X$ inclusive branching fraction is
unlikely to be explained by an enhanced $g^\star g \eta '$ form
factor, and an explanation outside the realm of the Standard Model may be needed to account for
this large rate.

\section{Acknowledgements}
We would like to thank A. Kagan and A. Ali for useful discussions
and for providing us with their calculations. We gratefully
acknowledge the effort of the CESR staff in providing us with
excellent luminosity and running conditions. D.~Cronin-Hennessy and
A.~Ryd thank the A.P.~Sloan Foundation. This work was supported by
the National Science Foundation, the U.S. Department of Energy, and
the Natural Sciences and Engineering Research Council of Canada.

\end{document}